# ETG Turbulence Induced Energy flux in the Large Laboratory Plasma

Prabhakar Srivastav[1, 2,], Rameswar Singh[3], L. M. Awasthi[1, 2,], A. K. Sanyasi[1], P. K. Srivastava[1], Ritesh Sugandhi[1, 2] and R. Singh [4]

[1]*Institute for Plasma Research, Gandhinagar 382428, India.*

[2]*Homi Bhabha National Institute, Mumbai 400085, India.*

[3]*University of California San Diego, United States.*

[4]*Advance Technology Center, NFRI, Rep. Korea.*

## Abstract

The Large Volume Plasma Device (LVPD) has successfully demonstrated excitation of Electron Temperature Gradient (ETG) driven turbulence in finite plasma beta ($\beta \sim 0.06 - 0.4$) condition , where the threshold condition for ETG turbulence is, $\eta_{\mathrm{ETG}} = L_n/L_T > 2/3$ satisfied, where, $L_n = \left[\frac{1}{n}\frac{dn}{dx}\right]^{-1}$ is density scale length and $L_{T_e} = \left[\frac{1}{T_e}\frac{dT_e}{dx}\right]^{-1}$ is temperature scale lengths [Mattoo et al,[1]]. The observed mode follows wave-vector scaling and frequency ordering as $k_\perp \rho_e \leq 1 << k_\perp \rho_i, \Omega_i < \omega << \Omega_e$, where $k_\perp$ is the perpendicular wave vector, $\rho_e$ , $\rho_i$ are Larmor radii of electron and ion, respectively, and $\Omega_i, \Omega_e$ $\omega$ are the ion, electron gyro frequencies and the mode frequency, respectively. Simultaneous measurement of fluctuations in electron temperature, $\delta T_e \sim (10 - 30)\%$, plasma density, $\delta n_e \sim (5 - 12)\%$ and potential $\delta V_f \sim (1 - 10)\%$ are obtained. Strong negative correlation with correlation coefficients $C_{\delta n - \delta \phi} \sim -0.8$ and $C_{\delta T - \delta \phi} \sim -0.9$ are observed between density and potential and temperature and potential fluctuations, respectively. These correlated density, temperature and potential fluctuations lead to generation of turbulent heat flux. The measured heat flux is compared with theoretically estimated heat flux from ETG model equations. The experimental result shows that net heat flux is directed radially outward.

## I.   INTRODUCTION

Plasma confinement and control of plasma transport still remains a significant challenge towards achieving fusion power. Plasma confinement is affected by anomalous cross field transport resulting from turbulence developed by collective modes/ instabilities which arises due to gradients in density, temperature, magnetic fields, etc [2–4] which are natural consequences of finite size of the device. These gradients work as a free energy source and leads to the generation of different instabilities, enabling anomalous transport of plasma particles and energy. It is well known that during low confinement (L) mode operation both ion and electron thermal transport are anomalous in nature. In high confinement (H-mode) scenario, due to the strong $E \times B$ shearing in the internal transport barrier, ion heat transport becomes neoclassical. However electron thermal transport still remain anomalous because the $E \times B$ shearing is probably still not strong enough to suppress the electron scale turbulence[5]. Focus is thus shifted to the understanding of physics of anomalous electron heat transport across the confining magnetic field, envisaging its implications for ITER and advanced Tokamak discharges[6–8].

In the past, extensive work has been reported on measurements of micro instabilities driven turbulence because of their possible role in causing anomalous particle and energy transport in fusion devices[4,9–12]. Outcome from these investigations suggest that transport by ion scale turbulence is largely understood but the Electron Temperature Gradient (ETG) driven turbulence, which is considered presently a major source of anomalous electron heat transport in fusion devices is still not properly understood[13]. Available literature on numerical and theoretical approaches shows significant advancement in understanding on ETG turbulence and transport but experimental investigations provide no direct evidence of its existence in tokamaks. The reason for no direct measurement of ETG may be due to the extremely small scale length in high magnetic field environment of fusion devices ($k_\perp \rho_e \sim 1$ when$\rho_e \sim \mu m$). The ETG mode is a short wavelength (shorter than ion larmor radius), high frequency (higher than ion cyclotron frequency) mode, $k_\perp \rho_e \leq 1 \ll k_\perp \rho_i$, $\Omega_i < \omega \ll \Omega_e$ where $k_\perp$ is perpendicular wave-vector, $\rho_e$ and $\rho_i$ are the larmor radii of electron and ion, respectively, $\Omega_e$ ($eB/m_e$), $\Omega_i(eB/m_i)$ and $\omega$ are electron, ion gyro-frequencies and mode frequency respectively. Theoretical models for slab ETG predicts that ETG is a fast growing mode driven by electron temperature gradient with characteristic growth rate $\gamma \approx k_y \rho_e \left( c_e / L_{T_e} \right)$ where, $c_e$ is the electron thermal velocity. Past investigations does provide indirect evidences of its existence during auxiliary heating investigations, carried out in devices like Tore Supra,

JET[14], DIII-D etc. but direct measurement of it is shown only in NSTX, where Mazzucato et al.[15], have shown its successful excitation by using electron scattering diagnostics. In recent times, linear devices like CLM[16], LVPD[1] etc. have demonstrated results supporting existence of ETG turbulence. Different mechanisms are used to meet the threshold condition of ETG in these devices. In CLM, this is done by heating electrons using a multi grid arrangement, while in LVPD, cross field diffusion concept is used by using transverse magnetic field produced by large electron energy filter (EEF)[17] for producing plasma suitable for studying unambiguous ETG turbulence. The EEF makes target plasma devoid of energetic electrons. Presence of energetic electrons could have poisoned plasma and subsequently, led to the excitation of beam plasma instabilities. In LVPD, wave length of ETG mode is scaled up to ~ few mm scales, which can be easily measured by conventional probes.

Introduction of Electron Energy Filter (EEF) divides LVPD plasma into three distinct regions of Source, EEF and Target plasmas. Source region covers the plasma volume between the cathode and the first surface of EEF, the EEF region covers the region between its two surfaces while the target plasma region extends from EEF second surface to the end plate[18]. Unambiguous, identification of ETG turbulence is successfully demonstrated in core region of target plasma ( $x \leq 45$cm and for $z = (20 - 150)$ cm) [1,19]. The work on ETG turbulence suggests two possible responsible mechanisms for different plasma beta regimes. In low plasma beta condition viz., $\beta \leq 0.1$ the slab ETG mode is primarily driven by parallel compression of electron motion along the magnetic field. While in high plasma beta, $\beta \geq 0.1$, ETG mode becomes unstable by its coupling with whistler mode which is responsible of finite compressibility due to $\delta B_z$ and diamagnetic compressibility due to nonzero $\nabla_x B$ effect[20].

In the present work, thermal heat conductivity is measured in the background of ETG turbulence in the core plasma region of LVPD. A specially designed triple Langmuir probe for real time measurement of temperature fluctuations in pulsed plasma of LVPD is used. The estimated thermal flux is compared with the numerically obtained values from slab ETG model equations.

The rest of the paper is organized as follows: the experimental setup and diagnostics are discussed in section II. The experimental observations on fluctuations and characterization are described in section III. The observation of heat flux, its comparison with numerically obtained values and justification are given in section IV. Finally, the paper ends with summary and conclusions described in section V.

## II. EXPERIMETAL SETUP AND MEASUREMENT FOR FLUCTUATIONS

The experiments for energy flux measurement is carried out in target region of Large Volume Plasma Device (LVPD)[21]. The LVPD is a cylindrical device containing plasma within it by using a combination of radial and axial confinement schemes[21]. The radial confinement is provided by a set of 10 garlanded magnet coils producing axial magnetic field, $B_z \sim 6.2\ G$ along its length and axial confinement by a pair of cusped ( $\sim 4$ kG, surface field) at back and end plates. These plates are mounted behind the plasma source and the other axial end of plasma column. The plasma source contains 36 numbers of hairpin shaped tungsten filaments ( $dia \sim 0.5$mm, and $L = 180$ mm), distributed on the periphery of a rectangle of size $1300 cm \times 900 cm$ cusped back plate. The pulsed Argon plasma of duration, $\Delta t_{\text{discharge}} \sim 9.2 ms$ is produced by appliying a discharge voltage of -70 V between filament assembly and the anode (device). The plasma system is equipped with a varying aspect ratio rectangular solenoid system (EEF) inside the plasma volume that produces a strong transverse magnetic field with respect to axial magnetic field ($B_{EEF} \perp B_z$). EEF is installed at the axial center of device. The EEF is highly transparent (82%) [17] and allows preferential cross field diffusion of thermal low energy electrons from source to target region. Bulk electron temperature of the target plasma is solely determined by the plasma transport across the transverse magnetic field of the EEF which in turn is decided by the magnetic field strength of the EEF. The EEF is made up of 155 numbers of equally spaced turns and is divided into 19 set of independent coils.

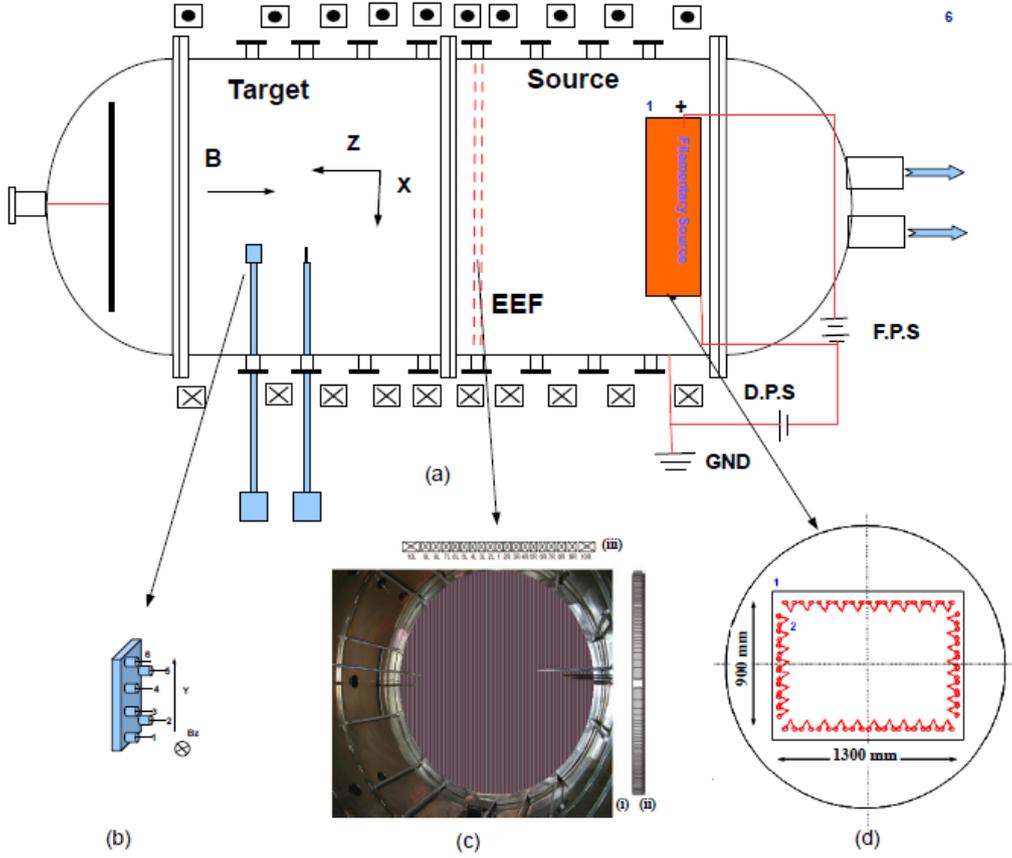

Fig 1: Schematic of experimental Device (Large Volume Plasma Device (LVPD)) (a) Top View of LVPD, (b) Langmuir probe assembly, (c) the photograph of EEF mounted within the device. The Top side bar in EEF photograph serves as coil identifier and RHS side bar defines extent of aspect ratios of each of the 19 coils, and (d) the front view of filament assembly in the source side of device.

The measurement of basic plasma parameters (electron temperature,$T_e$, plasma density, $n_e$, floating potential, $\varphi_f$ and plasma potential, $\varphi_p$) is carried out by using conventional cylindrical Langmuir probes of tungsten wire having $dia = 1$ mm and $L = 5$ mm and Centre Tapped Emissive Probes (CTEP)[22]. The plasma density is estimated from the ion saturation current, measured by keeping the probe at fixed bias of $\sim -80\,V$. Specially designed compensated Langmuir probes are used for the measurement of electron temperature by sweeping the probe between $-100V\ to +20V$ over a swept period of $500\mu s$[23]. All these probes are mounted at different axial locations in the ETG region on radially movable probe shafts[24]. The floating potential is measured in floating condition of Langmuir probe. The mean electron temperature obtained with single Langmuir probe (SLP) is compared with triple Langmuir probe (TLP) diagnostic[25]. Plasma potential is measured with use of centre tapped emissive probe (CTEP)[22]. The mean values of plasma parameters and fluctuations are obtained from the steady state plasma region of $(4 - 9.2)$ms from the onset of plasma discharge.

Specially configured TLP diagnostics is used for real time measurement of temperature and temperature fluctuations. The schematic shows the side view of probe assembly in 2(a) and the electrical configuration of TLP in 2(b).

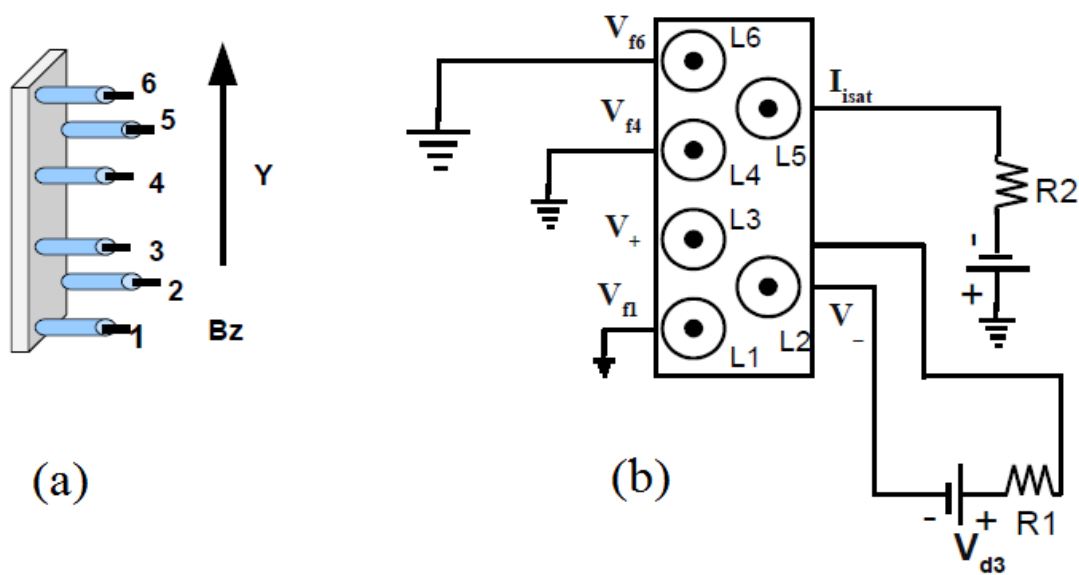

Fig 2: Schematic of 6- probe array for simultaneous measurement of particle and heat fluxes. Probes are arranged in the form of two consecutive triangles separated in 'Y'. The array is mounted on a motorized, radially movable linear probe drive. The side view of the probe array (a) and (b) is the electrical diagram for TLP diagnostics. The probe tips are approximately 30 mm away from the surface of ceramic holder.

The TLP diagnostic developed for real time, temperature fluctuation measurements in pulsed plasma of LVPD offer salient features namely, 1) bandwidth 300 kHz, 2) galvanic isolation of $250V$, 3) input impedance of $\gg 10M\Omega$ for voltage measurement and current measurement with shunt resistor $\sim 500\Omega$ respectively. The probe assembly shown in figure 2, consists of two sets of triple Langmuir probe assembly. First vertical array has 4 numbers of probes (L1, L3, L4, and L6) with separation,$\Delta y = 15$ mm, while second vertical array consists 2 probes (L2 & L4), vertically displaced, $\Delta y = 30$ mm. Probes numbering, L1, L2, L3 and L4 are used for electron temperature measurement. Probes are placed at different magnetic field lines and are transverse to the ambient magnetic field so as to avoid influence of shadowing effect. The TLP measurements are calibrated against single Langmuir probe measurements and TLP configuration is confirmed by obtaining I/V characteristics of double probe to ensure the symmetric current collection by the probe. The pair of L2 and L3 is used in double probe configuration and are powered by floating battery based power supply. It has characteristic features viz., variable voltage, negligible capacitance with respect to ground. The poloidally separated probes L1, L4 and L6 are used to measure floating potential. The Langmuir probe, L5 measures ion saturation current. By choosing suitable value of bias voltage between L2 and L3, $V_{23}$, such that $V_{23} \geq 3\,T_e/e$, we calculate electron temperature, $T_e$ by using expression $T_e = (V_+ - V_f)/\text{Log}\,(2)$, where $V_f$ is the average value of $V_{f1}$ and $V_{f4}$ .

The probe arrangement ensures that the parameters $V_f$, $I_{isat}$ are measured by probes placed in vicinity of each other to avoid the phase delay error. The fluctuating $\tilde{E}_\theta$ can be obtained by poloidally separated probes using $\tilde{E}_\theta = -\partial\delta\,\phi/\partial\,y$ and $\delta V_r$ is derived from $\delta E_\theta \times B$ drift. The radial velocity fluctuations are responsible for conductive ($n_o < \delta T_e \delta V_r >$) and convective heat flux ($T_e < \delta n_e \delta V_r >$) having correlation to temperature and density fluctuations respectively. The fluctuations in electron temperature, $\delta T_e$, density, $\delta n_e$ and potential, $\delta\phi$ are measured with a sampling rate of $1\,\text{MS}/s$. The data is acquired with a 12bit digitizer, PXI based data acquisition system. An ensemble of 100 shots from the steady state window is used for carrying out spectral analysis viz., correlation, coherency, phase, power spectra and joint wave number - frequency spectrum, $S(k,w)^{26,27}$.

## III.  EXPERIMETAL  OBSERVATIONS  AND  FLUCTUATION CHARCTERIZATION

Typical time profiles of plasma parameters in the target region for activated EEF are shown in fig. 3. The plasma discharge pulse (Fig. 3b) is accommodated within the pulse duration of the EEF current (Fig. 3a). The ion saturation current (Fig. 3c) do not show fluctuations in the early phase of the discharge but after $\sim 1ms$, fluctuations start appearing and are seen stabilizing after 6 ms from the onset of discharge. A similar trend in fluctuations is seen for other plasma parameters such as floating potential and electron temperature respectively. The floating potential, $V_f$ is measured using Langmuir probe at high impedance ($\geq 10\,M\Omega$) and is shown in fig 3(d). The evolution of mean electron temperature $T_e$ is recorded using triple Langmuir probe (Fig. 3e). The fluctuation and mean part of these plasma parameters are measured for the steady state plasma duration for further analysis to characterize the mean profile and nature of plasma turbulence.

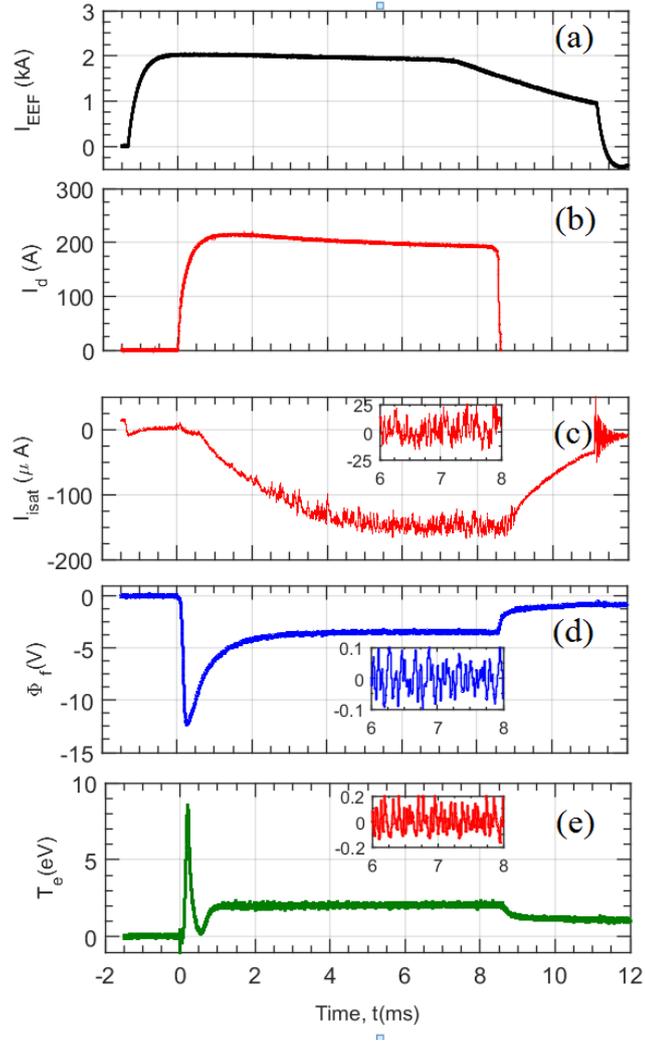

Fig 3: Time series traces of (*a*) filter current, $I_{\text{EEF}}$, (*b*) discharge current, $I_d$, (*c*) ion saturation current, $I_s$, (*d*) the floating potential, $\Phi_f$, and (*e*) the electron temperature, $T_e$ in the target plasma.

We revisited the ETG turbulence conditions and measured profiles of plasma density, and electron temperature in the target region of LVPD. The plasma potential is measured and found to be radially uniform in the target region, thus produces, $E_x \approx 0$ [28]. Radial profiles of plasma density and electron temperature are shown in fig 4a and 4b, respectively. The obtained $S(k, w)$ (Fig. 4c) shows the turbulent power corresponding to wave number $k_y$ (0.1 − 0.5) cm$^{-1}$ and frequency, $f (2 − 15) \, kHz$, which satisfies wavelength condition $k_\perp \rho_e \leq 1$ and $k_\perp \rho_i > 1$, and frequency ordering $\omega_{ci} < \omega \ll \omega_{ce}$ where $\omega_{ci}$ ( $\approx 1 \, krad/s$ ) is ion-cyclotron frequency, $\omega_{ce} (100 \, Mrad/s)$ is electron cyclotron frequency and $\omega$ (25 − 90 $krad/s$) is mode frequency, $\rho_e$ ($\sim 5 \, mm$) and $\rho_i$ ($\sim 40 \, cm$) are electron and ion Larmor radii, respectively and hence suggest that the turbulence is of electron scales. The scale length of density, $L_n = \left(\frac{1}{n}\frac{dn}{dx}\right)^{-1} \approx 300cm$ and electron temperature, $L_{T_e} = \left(\frac{1}{T_e}\frac{dT_e}{dx}\right)^{-1} \approx 55cm$ satisfies the threshold, $\eta_{ETG} = L_n/L_T > 2/3$ of ETG turbulence in the core region ($x \leq 50$cm)[1].

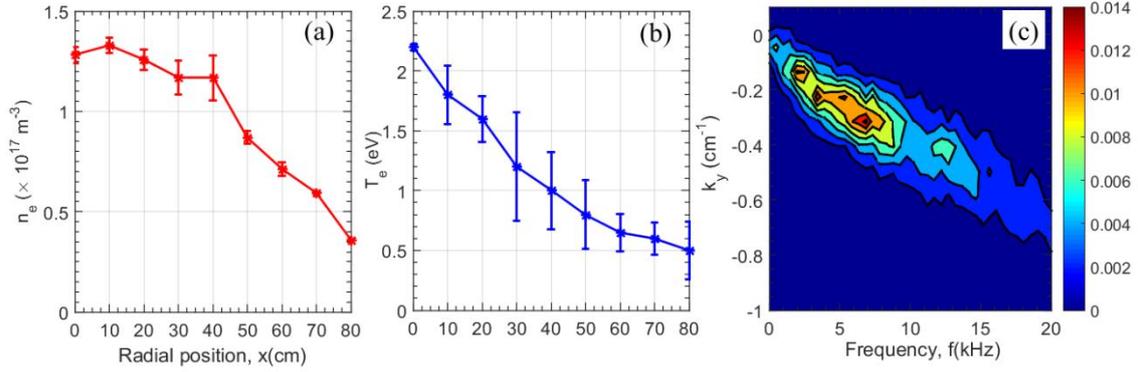

Fig 4: Radial profiles of (a) plasma density, $n_e$ , (b) electron temperature, $T_e$, and (c) the joint wave number –frequency, $S(k, f)$ , where, $k_y$ is the poloidal wave vector.

The time profiles of raw data fluctuations for each parameter is shown in figure 5 in steady state duration at radial position of $x = 20\ cm$. Figure 5a shows the time profile for temperature fluctuations, potential fluctuations and ion saturation current fluctuations are shown in figure 5b and 5c, respectively. This show in- phase correlation between density and temperature fluctuations and both are found out of phase to the potential fluctuations.

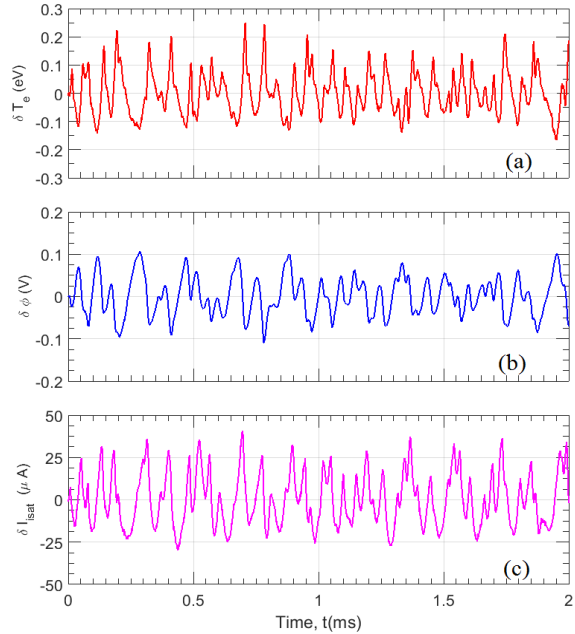

Fig 5: Typical raw data of fluctuation for steady state plasma of  (a) electron temperature, $\delta T_e$ (b) floating potential, $\delta\phi$ and (c) ion-saturation, $\delta I_{sat}$ at $x = 20\ cm$.

The cross correlation, $C(\tau)$ of potential fluctuation with density and temperature fluctuations is shown in the ETG dominated region. Obtained cross correlation between the said fluctuating paramters is found to be highly negative with correlation coefficients $C_{\delta n_e - \delta \phi} \sim -0.9$ and $C_{\delta T - \delta \phi} \sim -0.8$, respectively.

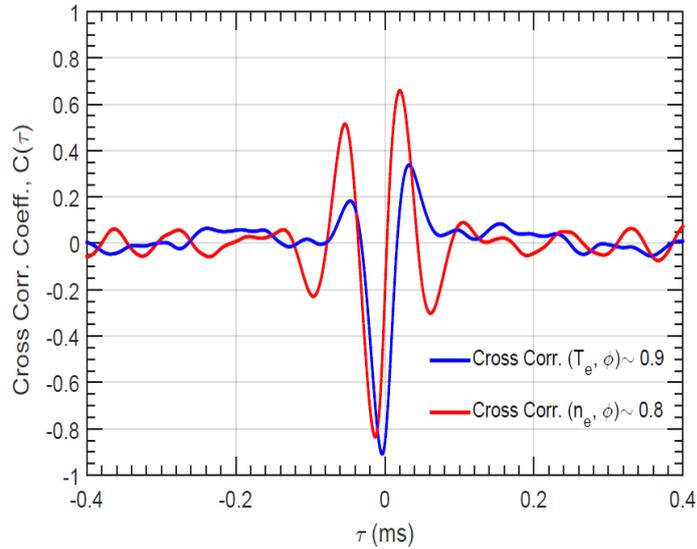

Fig 6: Cross correlation between potential with temperature and density fluctuations is shown. Potential fluctuation is found out of phase to both temperature fluctuations and density fluctuation. The measurement is taken at $x = 20 \ cm$.

The typical radial profiles of fluctuations in electron temperature, potential and ion saturation current are shown in Fig 7. The fluctuation levels varies for electron temperature from 2% to 20%, potential from 1% to 10% and ion saturation current fluctuations, which resembles plasma density between 2% - 8% from the plasma core to $x = 50cm$.

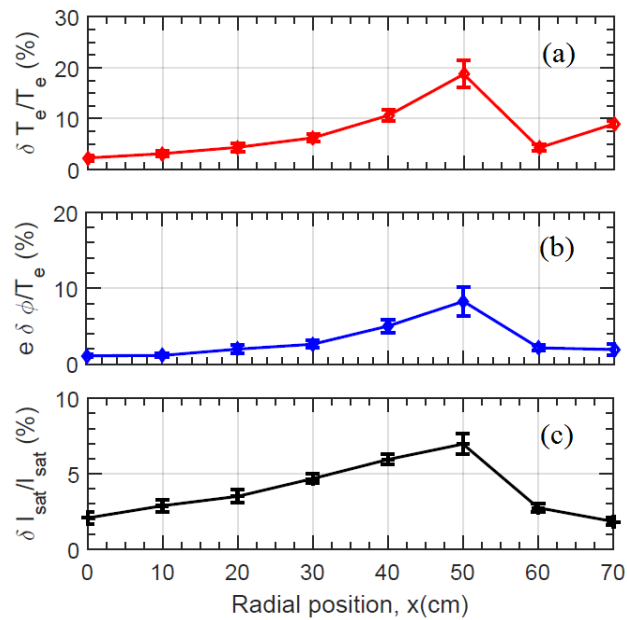

Fig 7: Radial profile of normalized fluctuations in (a) electron temperature, $\delta\,T_e/T_e$, (b) floating potential, $e\delta\phi/T_e$ and (c) ion saturation current, $\delta\,I_{sat}/I_{sat}$.

The spatiotemporal characteristics of the instability can be better envisaged by the coherency phase angle and power spectra of the instability. The auto-power, phase angle and coherency spectra are determined following the procedure described by Beall et al[26] and are shown in figure 8(a-c). The mode frequency has a broadband in the range $1-20\ kHz$, and the phase angle between the temperature and potential fluctuations is $\sim -150^0$. The coherency is significant (>0.8) up to $20\ kHz$ between the temperature and potential fluctuations.

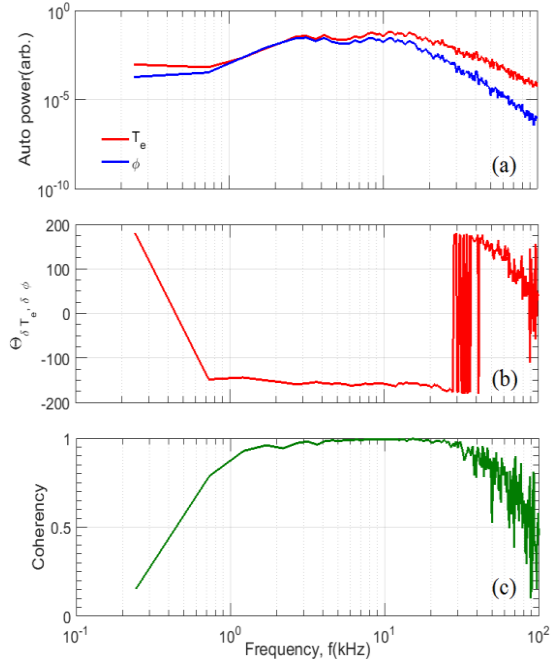

Fig8: The auto power (a), phase angle variation (b), and coherency (c) plot of temperature and potential fluctuations are shown. The spectrum has broadband nature with good coherency between $1-20\ kHz$. The measurements are carried out at $x=20\ cm$.

This observed phase angle between temperature and potential fluctuations is in good agreement of obtained values from ETG model equations with considerations of ion non-adiabatic response which is better explained in the section IV.

## IV. THEORETICAL MODEL AND MEASUREMENT OF HEAT FLUX

A theoretical expression is formulated for heat flux due to ETG scale fluctuations to compare it with experimental measurement of heat flux. The experimental heat flux is calculated with correlated measurement of potential and electron temperature fluctuations.

In LVPD, plasma beta is high hence both electrostatic (ES) and electromagnetic (EM) fluctuations are observed in ETG background. Both ES and EM fluctuations can contribute for total heat flux in the form of convective and conductive heat fluxes. The detailed contributions from both fluxes are explained herewith.

## A. Electrostatic Heat Flux

The electrostatic electron heat flux is defined as $q_e = \frac{3}{2} < \delta v_r \delta p_e > = \frac{3}{2} n_0 < \delta v_r \delta T_e > + \frac{3}{2} T_{e0} < \delta v_r \delta n_e >$ where the first and second terms are called conductive and convective heat fluxes.

Recall the electron temperature perturbation equation drift reduce for ETG scale

$$\frac{\partial}{\partial t}(T_e - \frac{2}{3} n_e) + (\eta_e - \frac{2}{3})\nabla_y \varphi + [\varphi, T_e - \frac{2}{3} n_e] = 0$$

which gives the linearized temperature fluctuation as

$T_{ek} = (\eta_e - \frac{2}{3}) \frac{k_y}{\omega} \varphi_k + \frac{2}{3} n_{ek}$ --------------------------------(1)

In dimensional form it reads

$\frac{\delta T_{ek}}{T_{e0}} = (\eta_e - \frac{2}{3}) \frac{c_e k_y \rho_e}{L_n \omega} \frac{e \delta \phi_k}{T_{e0}} + \frac{2}{3} \frac{\delta n_{ek}}{n_0}$-------------------(2)

Using $\frac{\delta n_{ek}}{n_0} = \frac{\delta n_{ik}}{n_0} = -\tau_e (A_k + i\delta_k) \frac{e \delta \phi}{T_{e0}}$ where $A_k$ is even in k and $\delta_k$ is odd in k, $\tau_e = T_e/T_i$ ratio of electron temperature to ion temperature.

As, $\delta_k = \frac{\sqrt{\pi}\omega}{k_\perp v_{thi}} \exp - \frac{\omega^2}{k_\perp^2 v_{thi}^2}$

Now using $\delta v_r = -c_e \rho_e \frac{\partial}{\partial y} \frac{e \delta \phi}{T_{e0}}$ for radial velocity fluctuation yields ($\Re$ stands for real part in the following)

$$q_{cond,es} = \frac{3}{2} n_o < \delta v_r \delta T_e >$$

$$= \frac{3}{2} n_o c_e T_{eo} \Re \sum_k \left[ \left(\eta_e - \frac{2}{3}\right) \frac{c_e \gamma_k (k_y \rho_e)^2}{L_n |\omega_k|^2} \left| \frac{e \delta \phi_k}{T_{eo}} \right|^2 + \frac{2}{3} \frac{\delta n_{ek}}{n_0} i k_y \rho_e \frac{e \delta \phi_k}{T_{eo}} \right]$$

$$= \frac{3}{2} n_o c_e T_{eo} \sum_k \left[ \left(\eta_e - \frac{2}{3}\right) \frac{c_e \gamma_k (k_y \rho_e)^2}{L_n |\omega_k|^2} \left| \frac{e \delta \phi_k}{T_{eo}} \right|^2 \right] + T_{eo} \, \Gamma_e$$----------------------------------(3)

Where,

$$\Gamma_e = \sum_k \pi^{\frac{1}{2}} \tau_e n c_e k_y \rho_e \left(\frac{\omega_r}{k_\perp V_{thi}}\right) \exp\left(-\frac{\omega_r^2}{k_\perp^2 V_{thi}^2}\right) \left|\frac{e\delta\phi}{T_{eo}}\right|^2$$

Clearly the usual conductive part also contains a distinct convective flux. Hence we subtract the convective part from above expression and define the conductive flux as

$$q_{cond} = \frac{3}{2} n_o < \delta v_r \delta T_e > - T_{eo} \, \Gamma_e$$

The total heat flux due to electrostatic fluctuations

$$q_{es} = q_{es}^{cond} + q_{em}^{conv} =$$

$$= \frac{3}{2} n_o c_e T_{eo} \sum_k \left[ \left(\eta_e - \frac{2}{3}\right) \frac{c_e \gamma_k (k_y \rho_e)^2}{L_n |\omega_k|^2} \left|\frac{e\delta\phi_k}{T_{eo}}\right|^2 \right] + T_{eo} \, \Gamma_e + \frac{3}{2} T_{e0} < \delta v_r \delta n_e >$$

$$= \frac{3}{2} n_o c_e T_{eo} \sum_k \left[ \left(\eta_e - \frac{2}{3}\right) \frac{c_e \gamma_k (k_y \rho_e)^2}{L_n |\omega_k|^2} \left|\frac{e\delta\phi_k}{T_{eo}}\right|^2 \right] + \frac{5}{2} T_{eo} \, \Gamma_e \qquad\qquad -----(4)$$

**Electromagnetic heat flux**

The electromagnetic heat flux in radial direction arises due to projection of parallel heat flux in radial direction due to fluctuations i.e. $q_{em} = < q_\parallel \delta B_r > / B_z$. A general expression for electromagnetic heat flux can be written as

$$q_{em} = \frac{3}{2} \frac{1}{B} < T n V_{\parallel e} B_r > = -\frac{3}{2} \frac{1}{eB} < T J_{\parallel e} B_r >$$

$$= -\frac{3}{2} \frac{T_{eo}}{eB} < \delta J_{\parallel e} \delta B_r > - \frac{3}{2} \frac{T_{eo}}{eB} J_{\parallel eo} < \frac{\delta T_e}{T_{eo}} \delta B_r >$$

$$= \frac{3}{2} T_{eo} \Gamma_e^{em} - \frac{3}{2} \frac{T_{eo}}{eB} J_{\parallel eo} < \frac{\delta T_e}{T_{eo}} \delta B_r >$$

The first term is known and considered as convective heat flux due to EM flux. The second term is conductive heat flux due to electromagnetic fluctuations and its contribution vanishes and this can be understood in the following manner;

$$< \frac{\delta T_e}{T_{eo}} \delta B_r > = -\Sigma_k i k_y R_A^* R_{T_e} \left|\frac{e\delta\phi_k}{T_{eo}}\right|^2$$

$$= \Sigma_k k_y \left[ (R_A^*)^r (R_{T_e})^i + (R_A^*)^i (R_{T_e})^r \right] \left|\frac{e\delta\phi_k}{T_{eo}}\right|$$

$$= 0$$

Where, $R_A = \dfrac{k_z \omega - \frac{5\tau_e}{3}\frac{\omega}{-}-\left(\eta_e-\frac{2}{3}\right)k_y}{\omega\left(\frac{\beta_e}{2}+k_\perp^2\right)\omega-\frac{\beta_e}{2}Kk_y}$ and $R_{Te} = \left[\left(\eta_e-\frac{2}{3}\right)\dfrac{c_e k_y \rho_e}{L_n \omega} - \dfrac{2}{3}\tau_e\left(1+i\dfrac{\pi^{\frac{1}{2}}\omega}{k_\perp V_{thi}}\exp\left(-\dfrac{\omega^2}{k_\perp^2 V_{thi}^2}\right)\right)\right]$

are electromagnetic and temperature response functions respectively. The above expression vanishes due to k space symmetry properties of $R_A$ and $R_T$. Hence the only surviving electromagnetic flux in the ETG turbulence is due to the electromagnetic particle flux.

$q_{em} = \dfrac{3}{2}T_{eo}\Gamma_e^{em}$ 　　　　　　　　　　- ----------------------------(5)

Thus, the total heat flux can be expressed as

$Q = q_{es} + q_{em} = \dfrac{3}{2}n_o c_e T_{eo}\sum_k\left[\left(\eta_e-\frac{2}{3}\right)\dfrac{c_e \gamma_k (k_y \rho_e)^2}{L_n |\omega_k|^2}\left|\dfrac{e\delta\phi_k}{T_{eo}}\right|^2\right]+\dfrac{5}{2}T_{eo}\Gamma_e+\dfrac{3}{2}T_{eo}\Gamma_e^{em}$

The significant portion of heat flux comprises mainly of electrostatic component as compared to the electromagnetic contribution and is significantly small ( $\Gamma_{em}/\Gamma_{es} = 10^{-5}$ )**.**

We compared the experimentally measured phase angle between temperature and potential fluctuations and average values of the heat flux with the theoretical estimated values. Figure 9 shows the comparison of phase angles. The phase angle is derived from equation (2) by taking into account the ion non-adiabatic response. This can be expressed as;

$\widetilde{T}_e = \left[\left(\eta_e-\frac{2}{3}\right)\dfrac{c_e k_y \rho_e}{L_n \omega} - \dfrac{2}{3}\tau_e\left(1+i\dfrac{\pi^{\frac{1}{2}}\omega}{k_\perp V_{thi}}\exp(-\dfrac{\omega^2}{k_\perp^2 V_{thi}^2})\right)\right]\tilde{\phi}$…………………………(6)

The experimentally obtained values of $k_\perp$ ($k_\perp \approx k_y$) and $\omega$ are chosen, corresponding to the maximum power of the observed mode in order to estimate the phase angle between the temperature and potential fluctuations. A close agreement between the two is observed in the ETG dominated region ( $x \le 50cm$ ). Indirect confirmation of the validity of model equations is envisaged from the fact that in the non ETG region, where they does not hold good, a significant deviation is observed in the cross phase angle.

The turbulent heat flux, $q_{cond} = \dfrac{3}{2}n_o < \delta T_e \delta V_r >$ has been estimated from the real time fluctuations of temperature and potential respectively. The temperature fluctuations are measured using a triple Langmuir probe assembly whereas, radial velocity fluctuation, $\delta V_r$ is derived from the potential fluctuations by using a poloidally separated pair of Langmuir probes.

The comparison of observed heat flux (red color), with analytical estimates (green color) and numerical estimates (blue color) are shown in figure 10, respectively. The analytical values are estimated by using equation (3), considering the value of $k_\perp$ ($k_\perp \approx k_y$) and $\omega$ for the peak power of the mode directly from $S(k_\perp, \omega)$. The values of $L_n, \eta_e$, $\rho_e$ and $c_e$ for this calculation are derived from the mean equilibrium profiles. An over estimation in the derived values for heat flux cannot be ruled out because of the finite spread in $k_\perp$ and $\omega$ values for the observed mode in $s(k_\perp, \omega)$ plot. The amplitude of potential fluctuations is directly taken from the experimental observations.

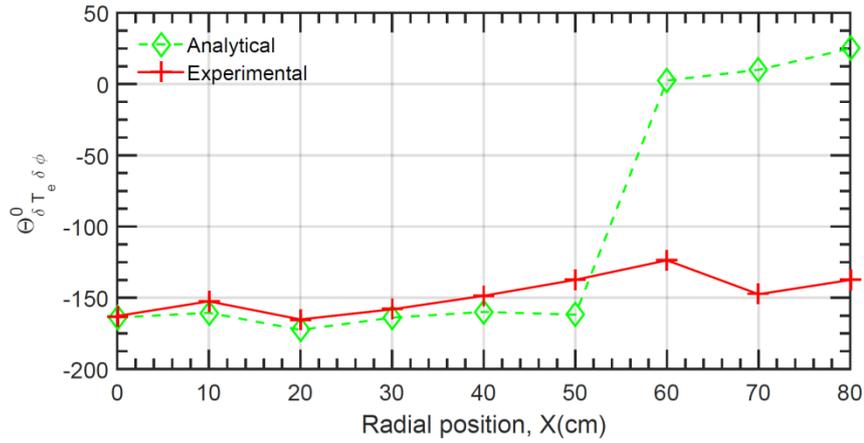

Fig 9: The comparison of experimentally obtained phase angle between temperature and potential fluctuations

Another numerical estimate of heat flux is obtained in the following way. Frequency and growth rates obtained from local W-ETG dispersion relation[1] for experimentally observed wave numbers and local mixing length estimate of intensity fluctuations are used to obtain numerical estimates of local flux. The comparison plot shows a good agreement with experimental observation of heat flux in the ETG dominated core region ($x \leq 50\ cm$) with analytical and numerical estimates.

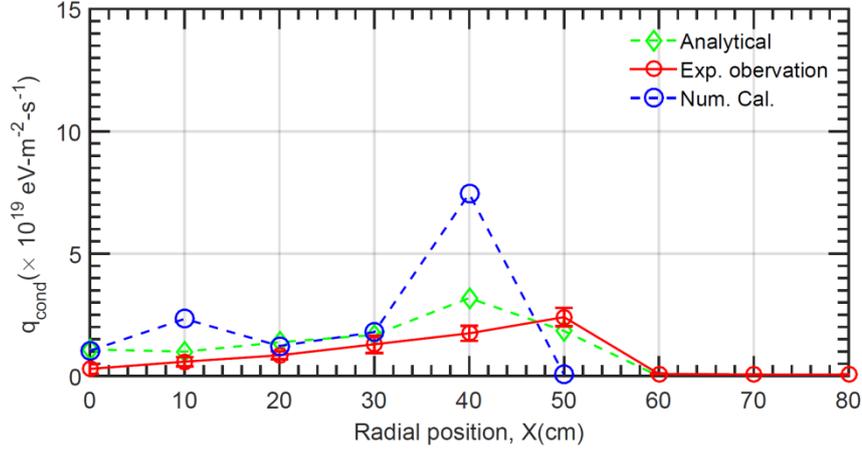

Fig 10: Comparison plot of conductive heat flux, $q_{cond}$ for experimental (red) measurement with analytical (green) and numerically (blue) estimated values for W-ETG turbulence.

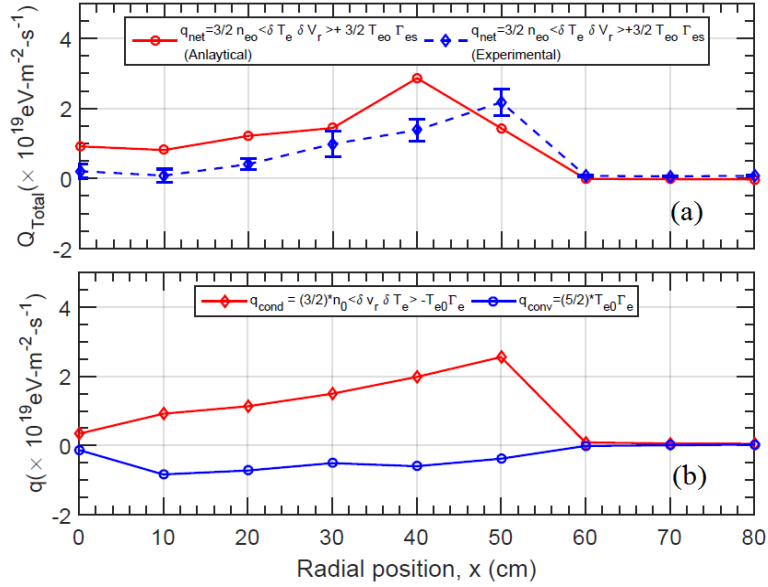

Fig 11. (a) Radial variation of total heat fluxes and (b) comparison of convective heat fluxes.

The total heat flux which is the sum of conductive and convective heat fluxes are shown in figure 11a. The total flux is found to be positive signifies that ETG driven turbulence can be responsible for heat loss. The figure 11b shows the conductive heat flux (red color) and convective heat flux (blue color) which is estimated from the observed heat flux due to temperature and potential fluctuations and particle flux measurement. Here, one can notice that conductive heat flux measured experimentally also have convective part and which must be subtracted from measured part so that conductive heat flux, $q_{cond} = \frac{3}{2} n_{eo} < \delta T_e \delta V_r >$ $-T_{eo}\Gamma_e$. Similarly, the convective heat flux is determined by adding particle flux contribution in conductive part to particle flux part which is the reason, $q_{conv} = \frac{5}{2}T_{eo} < \delta n_e \delta V_r >$.

We estimate the electron thermal conductivity at different radial locations over the entire core region of ETG dominated plasma. The electron thermal conductivity is calculated directly from the estimation of the fluctuation induced electron thermal flux.

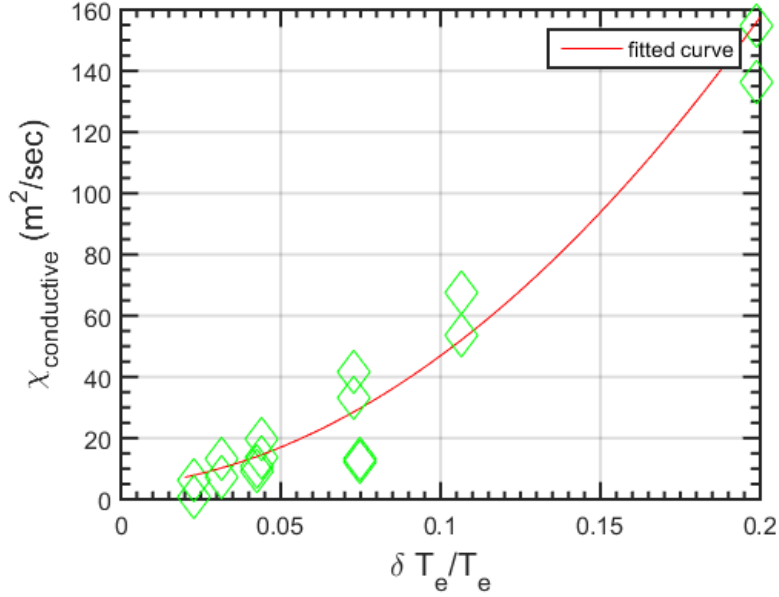

Fig 12: Electron thermal conductivity due to temperature fluctuations present in the system.

We attempted to calculate the variation of thermal conductivity with the level of temperature fluctuations in the core region. The heat conductivity is thus expressed in terms of total heat flux by the expression, $\chi = -\frac{Q}{n_{eo}\frac{dT_e}{dx}}$ where, $\chi$ is the heat conductivity, Q is the total heat flux and $\frac{dT_e}{dx}$ is the electron temperature gradient. Observation shows that thermal conductivity exhibits a quadratic dependency for normalized fluctuations between $5\% - 20\%$. This is expected from quasi-linear expressions of heat flux. Here, the temperature fluctuations are considered for different radial locations in the core region. The thermal conductivity maximizes at $x = 50\ cm$ where the temperature fluctuations have maximum amplitude as shown in figure 7.

## V.    SUMMARY AND CONCLUSION

In summary, we carried out measurement of radial profiles of electrostatic heat flux due to ETG turbulence in LVPD. Though the fluctuations in LVPD are electromagnetic in nature, the electromagnetic particle flux is smaller than the electrostatic flux. The electromagnetic particle flux is measured to be $10^{-5}$ times of the electrostatic flux. Simple analytical calculations of quasi-linear electromagnetic flux show that the electromagnetic conductive flux is zero. However, convective electromagnetic heat fluxes do exist due to electromagnetic particle flux. Hence, measurement of electrostatic heat flux is reported here.

Excitation of ETG turbulence is validated by measurements of power spectra, frequency scaling, cross phases between density - potential and temperature - potential. Correlation length, time and coherency are also measured to fully characterize turbulence. Theoretical estimates of frequency and cross phases are in close agreement with the respective experimental values. The cross angle between temperature and potential fluctuations differ from 180 degrees resulting in radially outward total heat flux. Conductive and convective heat fluxes are measured. Conductive heat flux is found to be radially outward and is larger than convective heat flux which is radially inward due to radially inward particle flux[29]. This signifies that total entropy of system is always positive and definite[30]. Thermal conductivity is found to scale quadratically with fluctuation intensity which is expected from quasi-linear estimates.

These laboratory observations may have significant implications for understanding the electron transport in fusion devices. Although present day tokamaks does not have high beta plasma but may have significance for the alternate magnetic concepts[31–33] as well as these results may be useful during sub- storm activities[34] taking place in magnetospheric plasmas as during this time, plasma beta is high.

## ACKNOWLEDGEMENT

We thank Dr Raju Daniel for reviewing the manuscript in IPR. Authors would like to express their sincere thanks to him for his critical evaluation of manuscript and for his useful suggestions towards improving the quality of the manuscript.